\pdfoutput=1

\documentclass[11pt,a4paper]{article}
\usepackage{bm}
\usepackage{jheppub}
\usepackage{mathrsfs}
\usepackage[T1]{fontenc}
\usepackage{rotating}
\usepackage{lscape}
\usepackage{bookmark}

\newcommand{\ket}[1]{\left|#1\right\rangle}      

\title{\boldmath On the Bethe states  of the one-dimensional supersymmetric $t-J$ model with generic open boundaries}

\author[a,b]{Pei Sun,}
\author[a,b]{Fakai Wen,}
\author[a,b]{Kun Hao,}
\author[c,d,e]{Junpeng Cao,}
\author[f]{Guang-Liang Li,}
\author[]{Tao Yang,${}^{a,b}$\footnote{Corresponding author}}
\author[a,b,g]{Wen-Li Yang, }
\author[a,b]{and Kangjie Shi}
\affiliation[a]{Institute of Modern Physics, Northwest University,
229 Taibai Beilu, Xian 710069, China} \affiliation[b]{Shaanxi Key Laboratory for
Theoretical Physics Frontiers, 229 Taibai Beilu, Xian 710069, China}
\affiliation[c]{Beijing National Laboratory for Condensed Matter
Physics, Institute of Physics, Chinese Academy of Sciences, 8 3rd South Street, Zhongguancun, Beijing
100190, China } \affiliation[d]{School of Physical Sciences,
University of Chinese Academy of Sciences, Beijing, China}
\affiliation[e]{Collaborative Innovation Center of Quantum Matter,
Beijing,  China} \affiliation[f]{Department of Applied Physics, Xian
Jiaotong University, 28 Xianning West Road, Xian 710049, China} \affiliation[g]{Beijing
Center for Mathematics and Information Interdisciplinary Sciences,
Beijing, 100048,  China} \emailAdd{sunpei@stumail.nwu.edu.cn}
\emailAdd{fakaiwen@126.com} \emailAdd{haoke72@163.com}
\emailAdd{junpengcao@iphy.ac.cn} \emailAdd{leegl@mail.xjtu.edu.cn}
\emailAdd{yangt@nwu.edu.cn}\emailAdd{wlyang@nwu.edu.cn} \emailAdd{kjshi@nwu.edu.cn}

\abstract{ By combining the  algebraic Bethe ansatz and the
off-diagonal Bethe ansatz, we investigate the supersymmetric $t-J$
model with generic open boundaries. The eigenvalues of the transfer
matrix are given in terms of an inhomogeneous $T-Q$ relation, and
the corresponding eigenstates are expressed in terms of  nested
Bethe states which have well-defined homogeneous limit.  This exact
solution provides basis for further analyzing the thermodynamic
properties and correlation functions of the model.}


\arxivnumber{1705.09478}

\begin{document}
\maketitle
\flushbottom

\section{Introduction}
\label{sec:introduction}

The $t-J$ model is one of the cornerstones in the study of
high-$T_{c}$ superconductivity~\cite{2-a-ctZhang}, which is a
large-$U$ limit of the single-band Hubbard
model~\cite{4-a-ctHu,5-a-ctEskes,7-a-ctHybertsen,8-a-ctHyberStech}.
The Hamiltonian of the model have played essential roles in
theoretical study of strongly correlated copperoxide based
materials~\cite{Sahinur2016}. In general, the Hamiltonian of the
supersymmetric $t-J$ model with the general boundary interaction terms is
given by
\begin{eqnarray}
H&=&-t\sum_{\alpha,j=1}^{L-1}\mathcal{P}\left[c_{j,\alpha}^{+}c_{j+1,\alpha}+c_{j+1,\alpha}^{+}c_{j,\alpha}\right]\mathcal{P}
+J\sum_{k=1}^{L-1}\left[\mathbf{S}_{k}\cdot \mathbf{S}_{k+1}-\frac{1}{4}n_{k}n_{k+1}\right]
+\sum_{l=1}^{L-1}n_{l}+n_{l+1}\nonumber\\
&&-\mu \hat{N}
+\xi_1n_{1}+2h_{1}^{z}S_1^z+2h_{1}^{-}S_1^{-}+2h_{1}^{+}S_1^{+}
+\xi_Ln_{L}+2h_{L}^{z}S_L^z+2h_{L}^{-}S_L^{-}+2h_{L}^{+}S_L^{+},
  \label{Hamilton}
\end{eqnarray}
where $t$ is the nearest neighbor hopping of electrons and $J$ is
the antiferomagetic exchange; $L$ is the total number of lattice
sites; The operators $c_{j,\sigma}$ and $c_{j,\sigma}^{+}$ are the
annihilation and creation operators of the electron with spin
$\sigma=\pm1$ on the lattice site $j$, which satisfies
anticommutation relations, i.e.,
$\{c_{i,\sigma}^{+},c_{j,\tau}\}=\delta_{i,j}\delta_{\sigma,\tau}$.
There are only three possible states at the lattice site $i$ due to
the factor $\mathcal{P}=(1-n_{j,-\sigma})$ ruled out double
occupancies; The operator $n_{j}=\sum_{\sigma=\pm}n_{j,\sigma}$
means the total number operator on site $j$ and
$n_{j,\sigma}=c_{j,\sigma}^{+}c_{j,\sigma}$; $\mu$ is the chemical
potential and $\hat{N}=\sum_{j=1}^{L}n_{j}$; $\xi_{1,L}$ are the
boundary chemical potentials; $h_{1,L}^{z}$ and $h_{1,L}^{\pm}$ are
the boundary fields; The spin operators
$S^{-}=\sum_{j=1}^{L}S_{j}^{-}$,\quad
$S^{+}=\sum_{j=1}^{L}S_{j}^{+}$ and $S^{z}=\sum_{j=1}^{L}S_{j}^{z}$ form the $su(2)$ algebra and can be expressed by
\begin{equation}
  S_{j}^{-}=c_{j,1}^{+}c_{j,-1},\qquad S_{j}^{+}=c_{j,-1}^{+}c_{j},\qquad S_{j}^{z}=\frac{1}{2}(n_{j,1}-n_{j,-1}).
\end{equation}

It is well-known that the one-dimensional $t-J$ model is integrable
at the supersymmetric point
$J=\pm2t$~\cite{35-a-ctKlai,36-a-ctsutherland,Sarkar}, and the model with the periodic boundary condition or the
diagonal boundaries has been studied by employing many Bethe ansatz
methods \cite{21-a-ctFoerster,22-a-ctGonzalez,23-a-ctEsslerJp,24-a-ctWangprl,25-a-ctFanhou,26-a-ctzhouyk,27-a-ctfanheng,
28-a-ctfanwadati,29-a-ctBedurftig,30-a-ctZnh,31-a-ctGalleas}. For
the non-diagonal boundary case, the nested algebraic Bethe ansatz method
doesn't work since the $U(1)$ symmetry is broken. With the help of
the off-diagonal Bethe ansatz~\cite{18-ctCaoYang,book Yang,
19-ctJPcao, 20-ctCaocui, 22-ctLiCao,23-ctJpcaoshi,
26-ctKhao}, the exact energy spectrum of the
one-dimensional supersymmetric $t-J$ model with unparallel boundary
fields has been obtained~\cite{32-a-ctZhangjp}. However, the
eigenstates (or Bethe states) which have played important roles in  applications of the model are still missing.

In this paper, we study the supersymmetric $t-J$ model with generic
integrable boundary conditions in grading: bosonic, fermionic and
fermionic (BFF). By combining  the graded nested algebraic Bethe ansatz
and off-diagonal Bethe ansatz, we obtain the Bethe states
which have well-defined homogeneous limit and
the corresponding eigenvalues of the transfer matrix of the model.
Numerical results for the small size systems suggest that the
spectrum obtained by the nested Bethe ansatz equations (BAEs) is
complete.

The paper is organized as follows. In section~\ref{sec:model}, the
associated graded $R$-matrix and corresponding generic integral
non-diagonal boundary reflection matrices are introduced. In
section~\ref{sec:AlgebraicBethe}, by using the graded algebraic
Bethe ansatz, we derive the eigenvalues of the transfer matrix of
the system which related with the eigenvalues of the nested transfer
matrix. In section~\ref{sec:ODBA}, the eigenvalues of the nested
transfer matrix are derived by off-diagonal Bethe ansatz, and the
Bethe states are also be given. In section~\ref{sec:T-Q}, we
construct the nested inhomogeneous $T-Q$ relation and the nested
Bethe ansatz equations of the supersymmetric $t-J$ model.
Section~\ref{sec:concluding} contains our results and give some
discussions.

\section{Integrability of the model}
\label{sec:model} \setcounter{equation}{0}

In this paper we consider
$J=2t=2$ which corresponds to the supersymmetric and integrable point \cite{37-a-ctEsslerKor}. The
integrability of the model is associated with the rational $R$-matrix $R(u)$ given by
\begin{equation}
\label{sun}
R_{12}(u)=\left(
  \begin{array}{ccc|ccc|ccc}
    u+\eta &  &  &  &  &  &  &  &  \\
     & u &  & \eta &  &  &  &  &  \\
     &  & u &  &  &  & \eta &  &  \\
    \hline
     & \eta &  & u &  &  &  &  &  \\
     &  &  &  & u-\eta &  &  &  &  \\
     &  &  &  &  & u &  & -\eta &  \\
    \hline
     &  & \eta &  &  &  & u &  &  \\
     &  &  &  &  & -\eta &  & u &  \\
     &  &  &  &  &  &  &  & u-\eta \\
  \end{array}
\right).
\end{equation}
The $R$-matrix $R(u)$ possesses the following properties
\begin{eqnarray}
\mbox{Initial condition:}&&R_{12}(0)=\eta P_{12}, \\
\mbox{Unitarity relation:}&&R_{12}(u)R_{21}(-u) = \rho_1(u)\,\times {\rm id},\\
\mbox{Crossing Unitarity relation:}&&R_{12}^{st_1}(-u+\eta)\,R_{21}^{st_1}(u)=\rho_2(u)\,\times {\rm id}.
\end{eqnarray}
Here $P_{12}$ is the graded permutation operator with the definition
\begin{equation}
  P_{\beta_{1}\beta_{2}}^{\alpha_{1}\alpha_{2}}=(-1)^{p(\alpha_{1})p(\alpha_{2})}\delta_{\alpha_{1}\beta_{2}}
  \delta_{\beta_{1}\alpha_{2}},
\end{equation}
$p(\alpha_i)$ is the Grassmann parities which is one for fermions
and zero for bosons. Here, we choose BFF grading which
means $p(1)=0,~ p(2)=p(3)=1$ and $R_{21}(u)=P_{12}R_{12}(u)P_{12}$,
$st_i$ denotes the super transposition in the $i$-th space
$(A^{st})_{ij}=A_{ji}(-1)^{p(i)[p(i)+p(j)]}$ and $ist_i$ denotes the
inverse super transposition. The functions $\rho_1(u)$ and
$\rho_2(u)$ are given by
\begin{eqnarray}
\rho_1(u)=-({u}-\eta)({u}+\eta), \quad
\rho_2(u)=-{u}({u}-\eta).
\end{eqnarray}
 Here and below we adopt the standard notations:
For any matrix $A\in {\rm End}({\rm\bf V})$, $A_j$ is an super
embedding operator in the ${Z_{2}}$ graded tensor space ${\rm\bf
V}\otimes {\rm\bf V}\otimes\cdots$, which acts as $A$ on the $j$-th
space and as identity on the other factor spaces. For $R\in {\rm
End}({\rm\bf V}\otimes {\rm\bf V})$, $R_{ij}$ is an super embedding
operator of $R$ in the ${Z_{2}}$ graded tensor space, which acts as
identity on the factor spaces except for the $i$-th and $j$-th ones.
The super tensor product of two operators
are defined through $(A\otimes B)_{\beta \delta}^{\alpha
\gamma}=(-1)^{[p(\alpha)+p(\beta)]p(\gamma)}A^{\alpha}_{\beta}B^{\gamma}_{\delta}$.
(For further details we refer the reader to
\cite{38-a-ctGrabinskiFrahm}).

The $R$-matrix is an even operator (i.e., the parities of the non-zero matrix elements $R_{bd}^{ac}$ of the
$R$-matrix  satisfies $p(a)+p(b)+p(c)+p(d)=0$) and satisfies the graded quantum Yang-Baxter equation (QYBE)
\begin{eqnarray}
R_{12}(u-v)\,R_{13}(u)\,R_{23}(v)=R_{23}(v)\,R_{13}(u)\,R_{12}(u-v).
\end{eqnarray}
In terms of the matrix entries, it reads
\begin{eqnarray}
\label{YBE-V}
  &&R(\lambda-u)_{\beta_{1}\beta_{2}}^{\alpha_{1}\alpha_{2}} R(\lambda)_{\gamma_{1}\beta_{3}}^{\beta_{1}\alpha_{3}}
  R(u)_{\gamma_{2}\gamma_{3}}^{\beta_{2}\beta_{3}}
  (-1)^{(p{(\beta_{1})}+p{(\gamma_{1})})p{(\beta_{2})}}\nonumber \\[1mm]
   && =R(u)_{\beta_{2}\beta_{3}}^{\alpha_{2}\alpha_{3}} R(\lambda)_{\beta_{1}\gamma_{3}}^{\alpha_{1}\beta_{3}}
   R(\lambda-u)_{\gamma_{1}\gamma_{2}}^{\beta_{1}\beta_{2}}
   (-1)^{(p{(\alpha_{1})}+p{(\beta_{1})})p{(\beta_{2})}}.
\end{eqnarray}
Let us now introduce the reflection matrix $K^-(u)$ and its dual one
$K^+(u)$. The former satisfies the graded reflection equation (RE)~\cite{Hengfan}
\begin{eqnarray}
 &&R_{12}(u_1-u_2)K^-_1(u_1)R_{21}(u_1+u_2)K^-_2(u_2)\nonumber\\
 &&=
K^-_2(u_2)R_{12}(u_1+u_2)K^-_1(u_1)R_{21}(u_1-u_2),\label{RE-V}
\end{eqnarray}
and the latter satisfies the dual RE which take the form \cite{39-a-ctAJbracken}
\begin{eqnarray}
&&R_{12}(u_2-u_1)K^+_1(u_1)\overset{\thickapprox}R_{21}(-u_1-u_2)^{ist_{1},st_{2}}K^+_2(u_2)\nonumber\\[1mm]
&&= K^+_2(u_2)\tilde{R}_{12}(-u_1-u_2)^{ist_{1},st_{2}}K^+_1(u_1)R_{21}(u_2-u_1), \label{DRE-V}
\end{eqnarray}
where
\begin{eqnarray}
  \overset{\thickapprox}R_{21}(u)^{ist_{1},st_{2}} &=& \left( \left[\{R_{21}^{-1}(u)\}^{ist_{2}}\right]
  ^{-1}\right)^{st_{2}},\\
  \tilde{R}_{12}(u)^{ist_{1},st_{2}} &=& \left( \left[\{R_{12}^{-1}(u)\}^{st_{1}}\right]
  ^{-1}\right)^{ist_{1}}.
\end{eqnarray}
For our case, the dual reflection equation (\ref{DRE-V}) reduces to
\begin{eqnarray}
&&R_{12}(u_2-u_1)K^{+}_1(u_1)R_{21}(-u_1-u_2+\eta)K^{+}_2(u_2)\nonumber\\[1mm]
&&=
K^{+}_2(u_2)R_{12}(-u_1-u_2+\eta)K^{+}_1(u_1)R_{21}(u_2-u_1).
\end{eqnarray}
In this paper we
consider the  generic non-diagonal $K$-matrices $K^-(u)$
 \begin{eqnarray}
 K^-(u)=\left(\begin{array}{ccc}\zeta+(2c-1)u&0&0\\
0&\zeta-u&2c_1u\\
0&2c_2u&\zeta+u\end{array}\right)\equiv \left(\begin{array}{ccc} k_{11}^- &0&0\\
0&k_{22}^-&k_{23}^-\\
0&k_{32}^-&k_{33}^- \end{array}\right). \label{K-matrix-1-RA}
\end{eqnarray}
Here the four boundary parameters
$c$, $c_1$, $c_2$ and $\zeta$ are not independent with each other, and  satisfy  a constraint
\begin{eqnarray}
c^2=c_1c_2+c. \nonumber
\end{eqnarray}
The dual non-diagonal reflection
matrix $K^+(u)$ is given by
\begin{eqnarray}
 K^+(u)=K^-(-u+\eta/2)\left|_{(\zeta,c,c_1,c_2)\rightarrow
(\zeta',c',c_1',c_2')}\right. \equiv \left(\begin{array}{ccc} k_{11}^+ &0&0\\
0&k_{22}^+&k_{23}^+\\
0&k_{32}^+&k_{33}^+ \end{array}\right),\label{K+-1}
\end{eqnarray}
with the constraint
\begin{eqnarray}
c'^2=c'_1c'_2+c'. \nonumber
\end{eqnarray}
In order to show the integrability of the system, we first
introduce the "row-to-row" monodromy matrices $T_0(u)$
and $\hat{T}_0(u)$
\begin{eqnarray}
T_0(u)&=&R_{0L}(u-\theta_L)R_{0\,L-1}(u-\theta_{L-1})\cdots
R_{01}(u-\theta_1),\\
\hat{T}_0(u)&=&R_{10}(u+\theta_1)R_{20}(u+\theta_{2})\cdots
R_{L0}(u+\theta_L),
\end{eqnarray}
where $\{\theta_j, j=1
\cdots L\}$ are the inhomogeneous parameters and $L$ is the number
of sites. The one-row monodromy matrices are the $3\times 3$
matrices in the auxillary space $0$ and their elements act on the
quantum space ${\rm\bf V}^{\otimes L}$. The tensor product is in the graded space, so
we can write
\begin{eqnarray}
  \left\{\left[T(u)\right]^{ab}\right\}_{\beta_{1}\dots\beta_{L}}^{\alpha_{1}\dots\alpha_{L}}
  &=&R_{0N}(u)^{a\alpha_{L}}_{c_{L}\beta_{L}} \dots R_{0j}(u)^{c_{j+1}\alpha_{j}}_{c_{j}\beta_{j}}
  \dots R_{01}(u)^{c_{2}\alpha_{1}}_{b\beta_{1}}\nonumber\\
  &&\times(-1)^{\sum_{j=2}^{L}(p(\alpha_{j})+p(\beta_{j}))\sum_{i=1}^{j-1}
  p(\alpha_{i})}.
\end{eqnarray}
For the system with open
boundaries, we need to define the double-row monodromy matrix
 \begin{eqnarray}
  \mathbb{T}_0(u)=T_0(u)K^-_0(u)\hat{T}_0(u),
\end{eqnarray}
which satisfies the similar relation as (\ref{RE-V}), in terms of matrix entries, they are
\begin{eqnarray}
  && R(u-\lambda)_{b_{1}b_{2}}^{a_{1}a_{2}}\mathbb{T}(u)_{c_{1}}^{b_{1}}R(u+\lambda)_{c_{2}d_{1}}^{b_{2}c_{1}}
  \mathbb{T}(\lambda)_{d_{2}}^{c_{2}}
  (-1)^{(p{(b_{1})}+p{(c_{1})})p{(b_{2})}}\nonumber\\[1mm]
  && =\mathbb{T}(\lambda)_{b_{2}}^{a_{2}}R(u+\lambda)_{b_{1}c_{2}}^{a_{1}b_{2}}\mathbb{T}(u)_{c_{1}}^{b_{1}}
  R(u-\lambda)_{d_{2}d_{1}}^{c_{2}c_{1}}
  (-1)^{(p{(b_{1})}+p{(c_{1})})p{(c_{2})}}.
  \label{mathTBA}
\end{eqnarray}
Then the transfer matrix of the system is constructed as
\begin{eqnarray}
t(u)=str_0\{K^+_0(u)\mathbb{T}_0(u)\}=\sum_{\alpha=1}^{3}(-1)^{p(\alpha)}\left[K^+_0(u)\mathbb{T}_0(u)\right]_{\alpha\alpha}
.\label{trans}
\end{eqnarray}
By using the \eqref{YBE-V}, \eqref{RE-V} and \eqref{DRE-V}, we can
prove the commutativity of $t(u)$. (For further details about the
commuting transfer matrix with boundaries for graded case, we refer
the reader to~\cite{39-a-ctAJbracken,40-a-ctGould,25-a-ctFanhou}).
The Hamiltonian (\ref{Hamilton}) can be constructed by taking the
derivative of the logarithm of the transfer matrix $t(u)$ of the system
\begin{eqnarray}
 H=-\frac{\eta}{2}\frac{\partial \ln t(u)}{\partial u}\Big|_{u=0,\{\theta_{j}=0\}}+\frac{\eta(2c-1)}{2\zeta}+\frac{\zeta'}{(c'-1/2)\eta-\zeta'}-\mu \hat{N}+L-1,
\end{eqnarray}
with the parameters chosen as follows:\newline
$\xi_1=-\frac{\eta}{2\zeta}(1-2c),\quad h_{1}^{z}=-\frac{\eta}{2\zeta} ,\quad h_{1}^{-}=-\frac{\eta}{2\zeta}c_2, \quad h_{1}^{+}=-\frac{\eta}{2\zeta}c_1,\quad
\xi_L=\frac{(c'-1/2)\eta}{(c'-1/2)\eta-\zeta'},\newline \quad h_{L}^{z}=\frac{-\eta/2}{(c'-1/2)\eta-\zeta'} ,\quad h_{L}^{-}=\frac{-\eta c_2'/2}{(c'-1/2)\eta-\zeta'} $ and $h_{L}^{+}=\frac{-\eta c_1'/2}{(c'-1/2)\eta-\zeta'}$.

\section{Nested algebraic Bethe ansatz}
\label{sec:AlgebraicBethe} \setcounter{equation}{0}
The block-diagonal structure of  the $K$-matrix (\ref{K-matrix-1-RA}) permits us to use the nested algebraic Bethe ansatz to construct the associated Bethe state and obtain the eigenvalue as follows. We first represent the double-row monodromy matrix $\mathbb{T}_0(u)=T_0(u)K^-_0(u)\hat{T}_0(u)$ in the form
\begin{eqnarray}
  \mathbb{T}_0(u)
   &=&\left(
        \begin{array}{ccc}
          A(u) & B_{1}(u) & B_{2}(u) \\
          C_{1}(u) & D_{11}(u) & D_{12}(u) \\
          C_{2}(u) & D_{21}(u) & D_{22}(u) \\
        \end{array}
      \right).\label{T-matrix}
\end{eqnarray}
Then the transfer matrix can be expressed by
\begin{eqnarray}
t(u)= \left[k_{11}^{+}(u)A(u)-\sum_{i,j=1}^{2} k_{i+1,j+1}^{+}(u)
D_{ji}(u)\right], \label{eigpa-RA}
\end{eqnarray}
where $k_{ij}^{\pm}$ is the $K^{\pm}$ matrix element in the $i$th
row and $j$th column.

Now we use the graded version of the nested algebraic Bethe ansatz method to obtain the
eigenvalues of the transfer matrix (\ref{eigpa-RA}). For this
purpose, we first define the reference state $|\Psi_0\rangle$ as
\begin{equation}
\label{psi0}
\ket{\Psi_{0}} = \bigotimes_{j=1}^{L}
\ket{0}_{j},\qquad \ket{0}_{j}=\left(\begin{array}{c}
              1 \\
              0 \\
              0
            \end{array}\right).
\end{equation}
From the relations \eqref{trans}, \eqref{T-matrix} and \eqref{psi0},
the elements of matrix $\mathbb{T}_0(u)$ acting on the reference state
$\ket{\Psi_0}$ give rise to
\begin{eqnarray}
A(u) \ket{\Psi_0} &=&k^{-}_{11}(u) a_{0}(u) \ket{\Psi_0}, \nonumber \\[1mm]
D_{11} (u) \ket{\Psi_0} &=& \left \{ \frac{\eta}{2u+\eta}k^{-}_{11}(u) a_{0}(u) + \left[ k^{-}_{22}(u) -  \frac{\eta}{2u+\eta}{k^{-}_{11}(u)} \right] b_{0}(u) \right\} \ket{\Psi_0}, \nonumber \\[1mm]
D_{22} (u) \ket{\Psi_0} &=& \left \{ \frac{\eta}{2u+\eta}k^{-}_{11}(u) a_{0}(u) + \left[ k^{-}_{33}(u) -  \frac{\eta}{2u+\eta}{k^{-}_{11}(u)} \right] b_{0}(u) \right\} \ket{\Psi_0}, \nonumber \\[1mm]
D_{12} (u) \ket{\Psi_0} &=& k^{-}_{23}(u)  b_{0}(u) \ket{\Psi_0},\nonumber\\[1mm]
D_{21} (u) \ket{\Psi_0} &=& k^{-}_{32}(u)  b_{0}(u) \ket{\Psi_0}, \nonumber \\[1mm]
B_{i} (u) \ket{\Psi_0} &\neq & 0, \quad C_{i}(u)\ket{\Psi_0}=0,\quad
i=1,2,
\end{eqnarray}
where
\begin{eqnarray}
b_0(u)=\prod_{j=1}^L(u-\theta_j)(u+\theta_j), \quad
a_0(u)=b_0(u+\eta). 
\end{eqnarray}
The operators $B_{1}(u)$ and $B_{2}(u)$ acting on the reference
state give nonzero values, and can be regarded as the creation
operators of the eigenstates of the system. Following the procedure
of the nested algebraic Bethe ansatz, the eigenstates of the
transfer matrix can be constructed  as
\begin{eqnarray}
\label{psi-RA}
|u_1,\ldots,u_M;\mathcal{F}\rangle
   = B_{a_{1}}(u_{1}) B_{a_{2}}(u_{2}) \dots B_{a_{M}}(u_{M})
\mathcal{F}^{a_{1} a_{2} \dots a_{M}} \ket{\Psi_0},
\end{eqnarray}
where we have used the convention that the repeated indices indict the sum over the values $1$,$2$, and
$\mathcal{F}^{a_{1}\dots a_{n}}$ is a function of the spectral parameters
$u_{j}$. Moreover,  the coefficients $\mathcal{F}^{a_{1}\dots a_{n}}$ are actually
the vector components of the nested Bethe state (see below (\ref{state F})). As the transfer matrix (\ref{eigpa-RA}) acting on the assumed
states (\ref{psi-RA}), we should exchange the positions
of the operators $A(u)$, $D_{ij}(u)$ and the operators
$B_{a_{j}}(u_j)$. With the help of the reflection equation
\eqref{mathTBA} and the Yang-Baxter equation (\ref{YBE-V}), we can derive
commutation relations
\begin{eqnarray}
B_{i}(u) B_{j}(v) &=&-B_{k}(v) B_{l}(u) \frac{r^{ij}_{lk}(u - v)}{u-v+\eta},\\[1mm]
A(u)B_{j}(v) &=& \frac{(u-v-\eta)(u+v)}{(u+v+\eta)(u-v)}B_{j}(v)A(u)
- \frac{\eta}{u+v+\eta}B_{i}(u)\tilde{D}_{ij}(v) \nonumber\\[1mm]
&&+\frac{2v\eta}{(u-v)(2v+\eta)}B_{j}(u)A(v),\label{AB-com} \\[1mm]
\tilde{D}_{ij} (u) B_{k}(v)
&=&\frac{r^{id}_{ef}(u+v+\eta)r^{fg}_{kj}(u-v)}{(u+v+\eta)(u-v)}
B_{d}(v)\tilde{D}_{eg}(u)+\eta\frac{r^{id}_{ej}(2u+\eta)}{(2u+\eta)(u-v)}B_{d}(u)
\tilde{D}_{ek}(v)\nonumber\\[1mm]
&&-\frac{2v\eta}{2u+\eta} \frac{r^{id}_{kj}(2u+\eta)}{(2v+\eta)(u+v+\eta)}B_{d}(u)A(v),
\label{ADB-com}
\end{eqnarray}
where $r_{ij}=u+\eta \mathbb{P}_{ij}$,
$\mathbb{P}_{\beta_{1}\beta_{2}}^{\alpha_{1}\alpha_{2}}=(-1)^{p(\alpha_{1})p(\beta_{2})}\delta_{\alpha_{1}\beta_{2}}
  \delta_{\beta_{1}\alpha_{2}}$ with the grading $p^{(1)}=p^{(2)}=1$, and
\begin{equation}
\tilde{D}_{ij}(u)= D_{ij}(u) -\delta_{ij}\frac{\eta}{2u+\eta} A(u).
\end{equation}
Acting the transfer matrix $t(u)$ on the state $\ket{\Psi}$ and
repeatedly using the commutation relations \eqref{AB-com} and
\eqref{ADB-com}, we obtain
\begin{eqnarray}
t(u)\,|u_1,\ldots,u_M;\mathcal{F}\rangle= \Lambda(u)\,|u_1,\ldots,u_M;\mathcal{F}\rangle + {\rm unwanted\; terms},
\end{eqnarray}
where the corresponding eigenvalue $\Lambda(u)$ is
\begin{eqnarray}
\Lambda(u) &=& \left[ -\frac{\eta}{2u+\eta} \sum_{i=2}^{3}
k_{ii}^{+}(u) +k_{11}^{+}(u) \right] k_{11}^{-}(u) a_{0}(u)
\prod_{i=1}^{M}
\frac{(u-u_{i}-\eta)}{(u-u_{i})} \frac{(u+u_{i})}{(u+u_{i}+\eta)} \nonumber \\
&&- b_{0}(u) \prod_{i=1}^{M} \frac{1}{(u-u_{i})(u+u_{i}+\eta)}
\hat{\Lambda}(u ,\{ u_{j} \}),\label{eigs}
\end{eqnarray}
and $\hat{\Lambda}\left(u,\{u_{j} \}\right)$ is the eigenvalue of the nested transfer matrix
$\hat{t}\left(u,\{ u_{j} \}\right)$ given by
\begin{eqnarray}
\hat{t}(u,\{ u_{j} \}) &=&
\frac{2u}{2u+\eta}\mbox{tr}_{\bar{0}} \left[ \bar{K}^{+}_{\bar{0}}
(u) r_{\bar{0} 1}(u+u_{1}+\eta) \dots r_{\bar{0}M}(u+u_{M}+\eta)
\right. \nonumber \\ &&\left. \times \bar{K}^{-}_{\bar{0}} (u) r_{M
\bar{0}}(u-u_{M}) \dots r_{1 \bar{0}}(u-u_{1}) \right], \label{Reduced-transfer-matrix-1}\\[1mm]
\bar{K}^{+}(u)&=&\left(\begin{array}{cc}
                  k^{+}_{22}(u) & k^{+}_{23}(u) \\[1mm]
                  k^{+}_{32}(u) & k^{+}_{33}(u)
                \end{array}\right),\label{k-first}
\\[1mm]
\bar{K}^{-}(u)&=&\frac{2u+\eta}{2u}\left(\begin{array}{cc}
                  k^{-}_{22}(u) - \frac{\eta}{2u+\eta}k^{-}_{11}(u) & k^{-}_{23}(u) \\[1mm]
                  k^{-}_{32}(u) & k^{-}_{33}(u) - \frac{\eta}{2u+\eta}k^{-}_{11}(u)
                \end{array}\right),\label{k-second}
\end{eqnarray}
namely,
\begin{eqnarray}
\hat{t}\left(u,\{ u_{j} \}\right) \ket{\mathcal{F}} &=&
\hat{\Lambda}\left(u,\{u_{j} \}\right) \ket{\mathcal{F}}.\label{Eigen-prob}
\end{eqnarray}
The vector components $\{F^{a_{1}\dots a_{n}}\}$ allow us to reconstruct the associated Bethe state
(\ref{psi-RA}), while the eigenvalue $\hat{\Lambda}\left(u,\{u_{j} \}\right)$ gives rise
to the associated eigenvalue (\ref{eigs}) of the transfer matrix $t(u)$ of the model.
We shall determine the eigenvalue $\hat{\Lambda}\left(u,\{u_{j} \}\right)$ and the corresponding eigenstate
$\ket{\mathcal{F}}$ in the next section. The condition that the unwanted
terms should be zero gives rise to  that the $M$ Bethe roots must satisfy
the associated  Bethe ansatz equations (BAEs)
\begin{eqnarray}
  1&=&\frac{K^{(1)}({u_{k}})a_0(u_{k})Q^{(1)}(u_{k}-\eta)}{(2u_{k}+\eta)b_0(u_{k})\hat{\Lambda}(u_{k} ,\{ u_{j} \})},\quad k=1\ldots M,\label{BAE-Open-1-1}
\end{eqnarray}
where
\begin{eqnarray}
Q^{(1)}(u)&=&\prod_{i=1}^{M}(u-u_i)(u+u_i+\eta),\\
K^{(1)}(u)&=&\left((2-4c')u^{2}+2\zeta'u-\eta\zeta'-\frac{1}{2}\eta^{2}+\eta^{2}c'\right)
\left(\zeta+(2c-1)u\right).
\end{eqnarray}
Some remarks are in order. It is easy to check that the nested Bethe state $|u_1,\ldots,u_M;\mathcal{F}\rangle $ given by (\ref{psi-RA})
and the eigenvalue
$\Lambda(u)$ given by (\ref{eigs}) both have well-defined homogeneous limit (i.e., $\theta_j\rightarrow 0$). This implies that in the homogeneous limit, the resulting Bethe states and the eigenvalue give rise to the eigenstate and the corresponding eigenvalue of the
super $t-J$ model described by the Hamiltonian (\ref{Hamilton}).

\section{Reduced spectrum problem}
\label{sec:ODBA}
\setcounter{equation}{0}

In the previous section, we have reduced searching eigenstates of the
original transfer matrix $t(u)$ (\ref{trans}) into the spectrum problem (\ref{Eigen-prob}) of
the nested transfer matrix $\hat{t}(u ,\{ u_{j} \})$ given by (\ref{Reduced-transfer-matrix-1}).
Now, we are in the position to calculate the eigenvalue $\hat{\Lambda}(u ,\{ u_{j} \})$ and the corresponding
eigenstate $\ket{\mathcal{F}}$ of the nested transfer matrix $\hat{t}(u ,\{ u_{j} \})$ which allows us to reconstruct
the Bethe state (\ref{psi-RA}) of the supersymmetric $t-J$ model. Because the
reflection matrices (\ref{k-first}) and (\ref{k-second}) have the
off-diagonal elements. The traditional algebraic Bethe ansatz is
invalid \cite{book Yang} due to the fact that the system doesnot have the obvious
reference state. Thanks to the works \cite{yang XXX, Nep13, Sam13, a-ct-zhang}, we can solve the spectrum problem (\ref{Eigen-prob}) as follows.
For simplicity, let $\lambda=u+\frac12\eta$ and
$\lambda_{j}=u_{j}+\frac12\eta$. We recognize the $\hat{t}(u ,\{
u_{j} \})$ as the transfer matrix of the open spin-$1/2$ XXX chain
of length $M$ with non-diagonal boundary terms. Following the
procedure in~\cite{yang XXX}
\begin{eqnarray}
\hat{t}\left(\lambda ,\{ \lambda_{j}\}\right)&=&\frac{2\lambda-\eta}{2\lambda}\bar {t}\left(\lambda ,\{ \lambda_{j}\}\right)\nonumber\\
&=&
\frac{2\lambda-\eta}{2\lambda}\mbox{Tr}_{\bar{0}}\left\{\bar{K}_{\bar{0}}^{+}(\lambda)\,
\bar{T}_{\bar{0}}(\lambda,\{{\lambda}_j\})\,
\bar{K}_{\bar{0}}^{-}(\lambda)\,
\widehat{\bar{T}}_{\bar{0}}(\lambda,\{{\lambda}_j\})\right\},
\label{nestedtransfer-RA}
\end{eqnarray}
where
\begin{eqnarray}
\bar{T}_{\bar{0}}(\lambda,\{{\lambda}_j\})
&=&r_{\bar{0} 1}(\lambda+\lambda_{1})
\dots r_{\bar{0}M}(\lambda+\lambda_{M}),\nonumber\\[1mm]
\widehat{\bar{T}}_{\bar{0}}(\lambda,\{{\lambda}_j\}) &=&r_{M
\bar{0}}(\lambda-\lambda_{M}) \dots r_{1
\bar{0}}(\lambda-\lambda_{1}), \label{offhatTmatric-RA} \\[1mm]
\bar{K}^{+}(\lambda)&=&\left(
                        \begin{array}{cc}
                        \zeta'+\lambda-\eta & 2c_{1}^{'}(-\lambda+\eta)\\[1mm]
                        2c_{2}^{'}(-\lambda+\eta)& \zeta'-\lambda+\eta \\
                        \end{array}
                      \right), \label{Kplus-RA2}\\[1mm]
\bar{K}^{-}(\lambda)&=&\left(
                      \begin{array}{cc}
                       -\lambda+1/2\eta+\zeta-c\eta & 2c_{1}\lambda  \\[1mm]
                        2c_{2}\lambda  & \lambda+1/2\eta+\zeta-c\eta \\
                      \end{array}
                    \right).\label{Kplus-RA}
\end{eqnarray}
We have checked  that \eqref{Kplus-RA} is the solution of the normal RE of the following form
\begin{eqnarray}
 &&r_{12}(u_1-u_2)\bar{K}^-_1(u_1)r_{21}(u_1+u_2)\bar{K}^-_2(u_2)\nonumber\\[1mm]
 &&=
\bar{K}^-_2(u_2)r_{12}(u_1+u_2)\bar{K}^-_1(u_1)r_{21}(u_1-u_2),
\end{eqnarray}
and that \eqref{Kplus-RA2} satisfies the
dual one. The $r$-matrix possesses the  properties
\begin{eqnarray}
\mbox{Initial condition:}&&r_{12}(0)=-\eta \bar{P}_{12},\\
\mbox{Unitary relation:}&&r_{12}(\lambda)r_{21}(-\lambda)=\bar{\rho}_1(\lambda)\,\times {\rm id},\\
\mbox{Crossing Unitary relation:}&&r_{12}^{t_1}(\lambda)\,r_{21}^{t_1}(-\lambda+2\eta)=\bar{\rho}_2(\lambda)\,\times {\rm id},\\
\mbox{PT-symmetry:}&&r_{21}(\lambda)=r^{t_1\,t_2}_{12}(\lambda).
\end{eqnarray}
Here $r_{21}(\lambda)=\bar{P}_{12}r_{12}(\lambda)\bar{P}_{12}$ and
$\bar{P}_{\beta_{1}\beta_{2}}^{\alpha_{1}\alpha_{2}}=\delta_{\alpha_{1}\beta_{2}}
  \delta_{\beta_{1}\alpha_{2}}$.
The functions $\bar{\rho_1}(\lambda)$ and $\bar{\rho_2}(\lambda)$ are given by
\begin{eqnarray}
\bar{\rho_1}(\lambda)=-({\lambda}-\eta)({\lambda}+\eta),\quad
\bar{\rho_2}(\lambda)=-{\lambda}({\lambda}-2\eta).
\end{eqnarray}
From the definition (\ref{nestedtransfer-RA}), we know that the
eigenvalue $\bar{\Lambda} (\lambda)$ of the transfer matrix
$\bar {t}\left(\lambda,\{ \lambda_{j}\}\right)$ is a polynomial of
${\lambda}$ and satisfies the  relations:
\begin{eqnarray}
\mbox{
Crossing
symmetry}:\bar{\Lambda}(\lambda)&=&\bar{\Lambda}(-\lambda+\eta),\label{Int-r-2-bar}\\
\mbox{ Asymptotic behavior}:\bar{\Lambda}(\lambda)&\sim&
{(-2-4c_{1}c_{2}'-4c_{1}'c_{2})u^{2M+2}},\quad
\lambda\rightarrow{\infty},\label{Int-r-2-2-bar}
\end{eqnarray}
and
\begin{eqnarray}
\bar{\Lambda}(\lambda_{j})\bar{\Lambda}(\lambda_{j}+\eta)=\frac{\triangle_{q}(\lambda_{j})}{{(\eta-2\lambda_{j})}
{(\eta+2\lambda_{j})}}, \quad j=1,\cdots, M, \label{Int-r-2-211-bar}
\end{eqnarray}
where
\begin{eqnarray}
\Delta_{q}(\lambda)&=&(2\eta+2\lambda)(2\eta-2\lambda)\nonumber\\
&&\times(\xi'^{2}-(1+4c_{1}'c_{2}')\lambda^{2})
\left((\xi+\frac{1}{2}\eta-c\eta)^{2}-(1+4c_{1}c_{2})\lambda^{2}\right)\nonumber\\
&&\times\prod_{j=1}^{M}(\lambda+\lambda_{j}-\eta)(\lambda-\lambda_{j}-\eta)(\lambda-\lambda_{j}+\eta)(\lambda+\lambda_{j}+\eta).
\end{eqnarray}
Some special points can also be calculated directly by using the properties of the $r$-matrix
and the reflection matrices $\bar{K}^{(\pm)}(u)$ as:
\begin{eqnarray}
&&\bar{\Lambda}(0)= \prod_{l=1}^M\bar{\rho_1}
(\lambda_l)tr \{\bar{K}^+(0)\}\bar{K}^-(0)\,\times{\rm id},\label{t1-1-bar} \\
&&\bar{\Lambda}(\eta)=\prod_{l=1}^M \bar{\rho_2}(\lambda_l+\eta)tr
\{\bar{K}^-(\eta)\}\bar{K}^+(\eta)\times {\rm id}. \label{t1-2-bar}
\end{eqnarray}
It is remarked that the above relations were derived independently  by the Separation of Variables \cite{Fra08}. These conditions (\ref{Int-r-2-bar})-(\ref{t1-2-bar}) allow us to construct the eigenvalue $\hat{\Lambda}(\lambda)$ in terms of an inhomogeneous $T-Q$ relation as \cite{yang XXX, Nep13}
\begin{eqnarray}
\hat{\Lambda}(\lambda)
&=&\frac{(2\lambda-\eta)}{2\lambda}\bar{\Lambda}(\lambda)\nonumber\\
&=&\frac{(2\lambda-2\eta)}{2\lambda}K^{(2)}(\lambda)\bar{a}(\lambda)
\frac{Q^{(2)}(\lambda+\eta)}{Q^{(2)}(\lambda)}
+ K^{(3)}(\lambda)\bar{d}(\lambda)
\frac{Q^{(2)}(\lambda-\eta)}{Q^{(2)}(\lambda)}\nonumber\\
 &&+(2\lambda-\eta)(2\lambda-2\eta)\bar{a}(\lambda)\bar{a}(-\lambda+\eta)
\frac{h}{Q^{(2)}(\lambda)},
\label{t2-3-3}
\end{eqnarray}
where
\begin{eqnarray}
\bar{a}(\lambda)&=&
\prod_{j=1}^M(\lambda+{\lambda}_j-\eta)(\lambda-{\lambda}_j-\eta),
\\
\bar{d}(\lambda)&=&
\prod_{j=1}^M(\lambda-{\lambda}_j)(\lambda+{\lambda}_j),\\
K^{(2)}(\lambda)&=&(-\sqrt{1+4(c'^{2}-c')}\lambda+\zeta')\nonumber\\
&&\times(\sqrt{1+4(c^{2}-c)}\lambda+\zeta+1/2\eta-c\eta),\\[2mm]
K^{(3)}(\lambda)&=&(\sqrt{1+4(c'^{2}-c')}(\lambda-\eta)+\zeta')\nonumber\\[1mm]
&&\times(\sqrt{1+4(c^{2}-c)}(-\lambda+\eta)+\zeta+1/2\eta-c\eta),\\[1mm]
h&=&\frac{1}{2}\left(-1-2(c_{1}'c_{2}+c_{2}'c_{1})+\sqrt{(1+4c_{1}'c_{2}')(1+4c_{1}c_{2})}\right),\\
Q^{(2)}(\lambda)&=&\prod_{j=1}^M (\lambda-w_j)(\lambda+w_j-\eta)\stackrel{{\rm def}}{=}\prod_{j=1}^M (\lambda-\nu_{j}-\frac{1}{2}\eta)(\lambda+\nu_{j}-\frac{1}{2}\eta).
\end{eqnarray}
Such parametrization obviously satisfies the crossing symmetry
\eqref{Int-r-2-bar},
asymptotic behavior (\ref{Int-r-2-2-bar}), production identity \eqref{Int-r-2-211-bar} and the values of the
special points (\ref{t1-1-bar}) and (\ref{t1-2-bar}). To ensure
$\hat{\Lambda}(\lambda)$ to be a polynomial, the residues of
$\hat{\Lambda}(\lambda)$ at the poles $w_{j}$ must vanish, i.e.,
the $M$ Bethe roots must satisfy the BAEs
\begin{eqnarray}
  (2w_{j}-2\eta)K^{(2)}(w_{j})\bar{a}(w_{j})Q^{(2)}(w_{j}+\eta)+2w_{j}K^{(3)}(w_{j})
  \bar{d}(w_{j})Q^{(2)}(w_{j}-\eta)\nonumber\\[1mm]
  +2w_{j}(2w_{j}-\eta)(2w_{j}-2\eta)\bar{a}(w_{j})\bar{a}(-w_{j}+\eta)h=0.\label{state BAES}
\end{eqnarray}
Now, we construct the eigenstates $|\mathcal{F}\rangle$ of the
nested transfer matrix $\hat{t}(\lambda)$. Following the ideas
in~\cite{Sam13, a-ct-zhang}, we first introduce two transformation matrices $g^{(\pm)}$:
\begin{eqnarray}
  g^{(-)} =\left(
              \begin{array}{cc}
                -1 & \frac{2c_{1}}{1-\sqrt{1+4c_{1}c_{2}}} \\
                1 & \frac{-2c_{1}}{1+\sqrt{1+4c_{1}c_{2}}} \\
              \end{array}
            \right),
\quad\quad
  g^{(+)} = \left(
            \begin{array}{cc}
              -m & \sqrt{1+mn}\!-\!1 \\
              -m & \!-\!\sqrt{1+mn}\!-\!1
            \end{array}
          \right),
\end{eqnarray}
where $m=\frac{-4c_{1}c_{2}-(2c_{1}c_{2}'-2c_{1}'c_{2})\sqrt{1+4c_{1}c_{2}}+2c_{1}c_{2}'+2c_{1}'c_{2}}
 {(1+2c_{1}c_{2}'+2c_{1}'c_{2})(\sqrt{1+4c_{1}c_{2}}+1)}$ and $n=\frac{-4c_{1}c_{2}-(2c_{1}'c_{2}-2c_{1}c_{2}')\sqrt{1+4c_{1}c_{2}}+2c_{1}'c_{2}+2c_{1}c_{2}'}
 {(1+2c_{1}c_{2}'+2c_{1}'c_{2})(\sqrt{1+4c_{1}c_{2}}-1)}$.
 The gauge matrices diagonalize the nested $K$-matrix  $\bar{K}^-(\lambda)$ given by (\ref{Kplus-RA})
and the matrix $g^{(-)}\bar{K}^{+}(\lambda)\,\{g^{(-)}\}^{-1}$ respectively, namely,
\begin{footnotesize}
\begin{eqnarray}
&&g^{(+)}\{g^{(-)}\bar{K}^+(\lambda)\{g^{(-)}\}^{-1}\}\{g^{(+)}\}^{-1}\nonumber\\
&&\!=\!-\frac{1+2c_{1}c_{2}'+2c_{1}'c_{2}}{\sqrt{1+4c_{1}c_{2}}}\left(
\begin{array}{cc}
\sqrt{1+mn}(\lambda-\eta)\!-\!\frac{\sqrt{1+4c_{1}c_{2}}}{1+2c_{1}c_{2}'+2c_{1}'c_{2}}\xi' & 0 \\
 0 & -\sqrt{1+mn}(\lambda-\eta)\!-\!\frac{\sqrt{1+4c_{1}c_{2}}}{1+2c_{1}c_{2}'+2c_{1}'c_{2}}\xi'\\
 \end{array}
 \right),\nonumber\\
&&g^{(-)}\bar{K}^-(\lambda)\{g^{(-)}\}^{-1}
=\left(
\begin{array}{cc}
 1/2\eta\!+\!\xi\!-\!c\eta\!+\!\lambda\sqrt{1+4c_{1}c_{2}}& 0 \\
 0 & 1/2\eta\!+\!\xi\!-\!c\eta\!-\!\lambda\sqrt{1+4c_{1}c_{2}} \\
 \end{array}
 \right).
\end{eqnarray}
\end{footnotesize}
With the gauge transformation, we can introduce the gauged monodromy matrix $\mathbb{U}(\lambda)$
\begin{eqnarray}
  \mathbb{U}(\lambda) = g^{(+)}\, \bar{T}(\lambda)\,\left\{g^{(-)}\bar{K}^{-}(\lambda)\{g^{(-)}\}^{-1}\right\}\,\hat{\bar{T}}(\lambda) \,\{g^{(+)}\}^{-1} = \left(
         \begin{array}{cc}
           \mathbb{A}(\lambda) & \mathbb{B}(\lambda) \\
           \mathbb{C}(\lambda) & \mathbb{D}(\lambda) \\
         \end{array}
       \right).\nonumber
\end{eqnarray}
Then it was shown in \cite{Sam13, a-ct-zhang}  that the eigenstate $|\mathcal{F}\rangle$ in (\ref{Eigen-prob}) can be expressed as
\begin{equation}\label{state F}
  |\mathcal{F}\rangle=\bigotimes_{j=1}^{M}\{g^{(-)}_{(j)}\}^{-1}\,
  \prod_{j=1}^{M}\mathbb{B}(w_{j})|0\rangle=\sum_{a_i=1,2} \mathcal{F}^{a_{1} a_{2} \dots a_{M}} |a_1,\dots,a_M\rangle,
\end{equation}
where the reference state $|0\rangle$ is
\begin{equation}
|0\rangle=\bigotimes_{j=1}^{M} \ket{1}_{j},\quad
\ket{1}_{j}=\left(\begin{array}{c}
              1 \\
              0
            \end{array}\right),
\end{equation}
provided that the parameters $\{w_j|j=1,\dots,M\}$  satisfy the BAEs (\ref{state BAES}). The corresponding vector components $\{\mathcal{F}^{a_{1} a_{2} \dots a_{M}}\}$
allow us to reconstruct the eigenstates  $|u_1,\ldots,u_M;\mathcal{F}\rangle$ given by (\ref{psi-RA}) of the original system \footnote{We have numerically checked, for small-site cases (such as $L=2,3$), that the states constructed by (\ref{psi-RA}) with vector components $\{\mathcal{F}^{a_{1} a_{2} \dots a_{M}}\}$ given by (\ref{state F}) give rise to the complete set of  eigenstates of the transfer matrix $t(u)$ given by (\ref{trans}), provided that the parameters $\{u_j\}$ and $\{v_j\}$ (or $\{w_j\}$) satisfy the BAEs (\ref{BAE-Open-3-1})-(\ref{BAE-Open-3-2}).  }.

\section{Nested inhomogeneous \texorpdfstring{$T-Q$}{} relation}
\label{sec:T-Q} \setcounter{equation}{0} Now we are ready to write
out the eigenvalues $\Lambda(u)$ of the transfer matrices $t(u)$  in  terms of some
inhomogeneous $T-Q$ relation with the help of \eqref{eigs} and \eqref{t2-3-3} as \footnote{Although the inhomogeneous $T-Q$ relation given by (\ref{3t-101}) is different from that obtained in \cite{32-a-ctZhangjp}, each of them gives rise to the complete set of eigenvalues of the transfer matrix. The $T-Q$ relation (\ref{3t-101}) takes advantage over one in \cite{32-a-ctZhangjp} is that it  leads to an simple form (\ref{state F})  of Bethe states  of the reduced spectrum problem (\ref{Eigen-prob}).}
\begin{eqnarray}
\Lambda(u)
&=&\frac{1}{(2u+\eta)} K^{(1)}(u)a_0(u)
\frac{Q^{(1)}(u-\eta)}{Q^{(1)}(u)}
\nonumber \\
&&-\frac{
(2u-\eta)}{(2u+\eta)}K^{(2)}(u+\frac{1}{2}\eta)b_0(u)\frac{Q^{(1)}(u-\eta)Q^{(2)}(u+\frac{3}{2}\eta)}{Q^{(1)}(u)Q^{(2)}(u+\frac{1}{2}\eta)}
\nonumber \\ &&-K^{(3)}(u+\frac{1}{2}\eta)b_0(u)
\frac{Q^{(2)}(u-\frac{1}{2}\eta)}{Q^{(2)}(u+\frac{1}{2}\eta)}-2u(2u-\eta)b_0(u)
\frac{hQ^{(1)}(u-\eta)}{Q^{(2)}(u+\frac{1}{2}\eta)}, \label{3t-101}
\end{eqnarray}
\begin{table}[t]
\centering 
\begin{scriptsize}
\begin{tabular}{|c ccc|c|c|c|c|} \hline
$u_{1}$ & $u_{2}$ & $\nu_{1}$ & $\nu_{2}$ & $E_{n}$ & $n$  \\\hline
$-0.1000-1.6602i$  &  $0.1004-0.0000i$  &  $0.1264+3.3108i$  &  $0.1264-3.3108i$ & $-5.312156$ & $1$  \\
 $-0.1000-0.2048i$  &  $0.1004+0.0000i$  &  $0.8095-3.4060i$  &  $0.8095+3.4060i$ & $-4.555656$ & $2$  \\
$0.1005+0.0000i$  & $-$ & $0.0000-3.3070i$ & $-$ & $-3.325040$ & $3$  \\
 $-0.1000-1.6539i$  &  $-0.1000-0.2053i$  &  $0.0255+3.3085i$  &  $0.0255-3.3085i$ & $-3.218186$ & $4$  \\
 $-0.1000-3.7095i$  &  $-0.1000-0.1000i$  &  $0.0000-4.6812i$  &  $0.1496-0.0000i$ & $-1.996355$ & $5$  \\
$-0.1000-2.3555i$  & $-$ &  $0.0000-3.7060i$ & $-$ & $-1.992804$ & $6$  \\
$-0.1000+0.2040i$  & $-$ &  $-0.0000+3.3148i$ & $-$ & $-1.225154$ & $7$  \\
$-$  & $-$ &  $-$ & $-$ & $0$ & $8$  \\
$-0.1000-0.0999i$  & $-$ &  $-0.1496-0.0000i$ & $-$ & $0.001822$ & $9$  \\
\hline \end{tabular}
\end{scriptsize}
\caption{\label{table:BAEs1} Solutions of BAEs
\eqref{BAE-Open-3-1} and \eqref{BAE-Open-3-2} where $L=2$ with the
parameters $\eta=0.2,\mu=2, \zeta=0.1, c=0.1, c_{1}=-0.5,
\zeta'=-0.5, c'=-0.3$ and $c_{1}'=-0.7$ for the case of $E_{n}$ is
the corresponding eigenenergy. The energy $E_{n}$ calculated from
\eqref{energy_E} is the same as that from the exact diagonalization
of the Hamiltonian (\ref{Hamilton}).}
\end{table}
\begin{landscape}
\begin{table}[t]
\centering 
\begin{scriptsize}
\begin{tabular}{|cccccc|c|c|} \hline
$u_1$ & $u_2$ & $u_3$ & $\nu_1$ &  $\nu_{2}$ & $\nu_{3}$ & $E_{n}$ &
$n$ \\\hline $0.1183-0.0000i$  &  $0.8870+0.0000i$  &
$0.4749-0.0000i$  &  $-0.0000+2.1856i$  & $1.5149-4.0404i$  &
$1.5149+4.0404i$ & $-7.229050$ & $1$ \\
 $0.1073-0.0000i$  &  $-0.1000-0.1218i$  &  $-0.1000+2.3587i$  &  $0.0000-0.1091i$
 & $0.9332-4.1573i$  &  $0.9332+4.1573i$ & $-5.595946$ & $2$  \\
$0.1183-0.0000i$  &  $0.4290-0.0000i$  & $-$ &  $0.8379-3.4029i$  &
$0.8379+3.4029i$ & $-$ & $-5.210221$ & $3$\\
 $1.4165-0.0000i$  &  $-0.3183-0.0000i$  & $-$ &  $0.9773-3.3657i$
 &$0.9773+3.3657i$ & $-$ & $-5.079400$ & $4$  \\
  $0.1182-0.0000i$  &  $0.4997-0.0000i$  &  $-0.1000-0.0701i$  &  $-0.0000+1.9912i$
  &$1.4407-4.0634i$  &  $1.4407+4.0634i$ & $-4.495822$ & $5$  \\
 $-0.1000+0.0701i$  &  $0.1183+0.0000i$  &  $0.8474+0.0000i$  &  $0.0000-2.1041i$
 &$1.4824-4.0500i$  &  $1.4824+4.0500i$ & $-4.426045$ & $6$  \\
 $0.0005-0.0704i$  &  $0.0005+0.0704i$  &  $-0.1000+2.3565i$  &  $-0.0000+0.0684i$
 & $0.9328-4.1576i$  &  $0.9328+4.1576i$ & $-4.253561$ & $7$  \\
 $1.4210+0.0000i$  &  $0.4273+0.0000i$  & $-$ &  $0.9950-3.3623i$
 & $0.9950+3.3623i$ & $-$ & $-4.166597$ & $8$  \\
 $-0.1000-0.1216i$  &  $0.1073+0.0000i$  & $-$ &  $0.0000-3.8739i$
 & $0.0000-0.1092i$ & $-$ & $-3.598898$ & $9$  \\
  $-0.1000+0.0701i$  &  $0.8813-0.0000i$  &  $0.4760+0.0000i$  &  $1.5100-4.0418i$
  &$1.5100+4.0418i$  &  $0.0000+2.1735i$ & $-3.484233$ & $10$  \\
   $0.1183+0.0000i$ & $-$ & $-$  &  $0.0000+3.3063i$
   &$-$ & $-$  & $-3.061679$ & $11$  \\
 $-0.1000-0.1733i$  &  $-0.1000-2.8910i$  & $-$ &  $-0.0000+4.5408i$
 &$-0.0000-0.0711i$ & $-$ & $-2.996075$ & $12$  \\
  $-0.1000-0.0601i$  &  $-0.1000+2.3432i$  &  $-0.1000-0.3120i$  &  $0.9298-4.1596i$
  &$0.9298+4.1596i$  &  $0.1995-0.0000i$ & $-2.682705$ & $13$  \\
 $-0.1000+0.0700i$  &  $0.1183+0.0000i$  & $-$ &  $0.8135-3.4056i$
 &$0.8135+3.4056i$ & $-$ & $-2.378366$ & $14$  \\
 $0.0004-0.0699i$  &  $0.0004+0.0699i$  & $-$ &  $0.0000-3.8760i$
 &$0.0000-0.0680i$ & $-$ & $-2.256786$ & $15$  \\
  $0.4178-0.0000i$ & $-$ & $-$  &  $0.0000-3.2852i$
  &$-$ & $-$  & $-2.154975$ & $16$  \\
 $1.7975-0.0000i$ & $-$ & $-$   &  $0.0000-2.9281i$
 &$-$ & $-$ & $-2.011140$ & $17$  \\
  $-0.1000-0.0577i$  &  $-0.1000+0.1731i$  &  $-0.1000+4.5434i$  &  $0.1682-0.0768i$
  &$0.1682+0.0768i$  &  $0.0000-5.7342i$ & $-1.997344$ & $18$  \\
 $-0.1000-0.0700i$  &  $-0.6290+0.0000i$  & $-$ &  $0.8334+3.4033i$
 &$0.8334-3.4033i$ & $-$ & $-1.464602$ & $19$  \\
 $-0.1000+0.0700i$  &  $1.4155+0.0000i$  & $-$ &  $0.9731-3.3664i$
 &$0.9731+3.3664i$ & $-$ & $-1.333802$ & $20$  \\
  $-0.1000+0.1730i$ & $-$ & $-$   &  $0.0000+0.0709i$
  &$-$ & $-$  & $-0.998177$ & $21$  \\
 $-0.1000-0.0577i$  &  $-0.1000+2.8833i$  & $-$ &  $0.1777+0.0000i$
 &$-0.0000-4.5376i$ & $-$ & $-0.995481$ & $22$  \\
 $-0.1000-0.0601i$  &  $-0.1000+0.3107i$  & $-$ &  $-0.0000-3.8889i$
 &$-0.1993-0.0000i$ & $-$ & $-0.686247$ & $23$  \\
 $-$  &  $-$  &  $-$  &  $-$
 &$-$  &  $-$ & $0$ & $24$  \\
 $-0.1000+0.1731i$  &  $-0.1000-0.0577i$  & $-$ &  $0.1682-0.0768i$
 &$0.1682+0.0768i$ & $-$ & $0.001770$ & $25$  \\
  $-0.1000+0.0700i$ & $-$ & $-$  &  $0.0000+3.3113i$
  & $-$ & $-$  & $0.684187$ & $26$  \\
 $-0.1000-0.0577i$ & $-$ & $-$  &  $0.1776+0.0000i$
 & $-$ & $-$ & $1.000607$ & $27$ \\
\hline \end{tabular} \caption{\label{table:BAEs} Solutions of BAEs
(\ref{BAE-Open-3-1}) and (\ref{BAE-Open-3-2}) where $L=3$ with the
parameters $\eta=0.2, \mu=2, \zeta=0.1, c=0.1, c_{1}=-0.5,
\zeta'=-0.5, c'=-0.3$ and $c_{1}'=-0.7$ for the case of $E_{n}$ is
the corresponding eigenenergy. The energy $E_{n}$ calculated from
(\ref{energy_E}) is the same as that from the exact diagonalization
of the Hamiltonian (\ref{Hamilton}).}
\end{scriptsize}
\end{table}
\end{landscape}

\noindent where the $2M$ Bethe roots must satisfy the BAEs
(\ref{BAE-Open-1-1}) and (\ref{state BAES}), namely,
\begin{eqnarray}
 1&+&\frac{(2\nu_{l}-\eta)}{(2\nu_{l}+\eta)}\frac{K^{(2)}(\nu_{l}+\frac{1}{2}\eta)}{K^{(3)}(\nu_{l}+\frac{1}{2}\eta)}
\frac{Q^{(1)}(\nu_{l}-\eta)Q^{(2)}(\nu_{l}+\frac{3}{2}\eta)}{Q^{(1)}(\nu_{l})Q^{(2)}(\nu_{l}-\frac{1}{2}\eta)}\nonumber\\[1mm]
&=&-h{(2\nu_{l})(2\nu_{l}-\eta)
}
\frac{Q^{(1)}(\nu_{l}-\eta)}
{K^{(3)}(\nu_{l}+\frac{1}{2}\eta)Q^{(2)}(\nu_{l}-\frac{1}{2}\eta)},\quad l=1\ldots M,\label{BAE-Open-3-1}\\[1mm]
1&=&(2u_k-\eta)\frac{K^{(2)}(u_{k}+\frac{1}{2}\eta)b_0(u_k)}{K^{(1)}({u_k})a_0(u_k)}
\frac{Q^{(2)}(u_k+\frac{3}{2}\eta)}{Q^{(2)}(u_k+\frac{1}{2}\eta)},\quad k=1\ldots M.\label{BAE-Open-3-2}
\end{eqnarray}

In the homogeneous limit, the corresponding $T-Q$ relation and associated BAEs become
(\ref{3t-101}) and (\ref{BAE-Open-3-1})-(\ref{BAE-Open-3-2}) by setting $\theta_j=0,\,j=1,\ldots,N$. Therefore
the energy of the Hamiltonian \eqref{Hamilton} reads
\begin{eqnarray}
E&=&-\frac{\eta}{2}\frac{\partial \ln \Lambda(u)}{\partial u}|_{u=0,\{\theta_j\}=0}+\frac{\eta(2c-1)}{2\zeta}+\frac{\zeta'}{(c'-1/2)\eta-\zeta'} -\mu M +L-1 \nonumber\\
&=&-\sum_{k=1}^{M}\frac{\eta^2}{u_k(u_k+\eta)}-\mu M,
\label{energy_E}
\end{eqnarray}
where the $2M$ parameters  $\{u_j|j=1,\ldots,M\}$ and $\{v_j|j=1,\ldots,M\}$  satisfy the resulting BAEs
\eqref{BAE-Open-3-1} and \eqref{BAE-Open-3-2}. Here we present the
results for the $L=2$ and $L=3$ cases: the numerical solutions of the
BAEs are shown in table~\ref{table:BAEs1} and
table~\ref{table:BAEs}, which indicate that the eigenvalues are identical with the
results we get from the exact diagonalization of the Hamiltonian
(\ref{Hamilton}). Numerical results for the small-site cases
suggest that the spectrum obtained by the nested BAEs (\ref{BAE-Open-3-1})-(\ref{BAE-Open-3-2}) is complete.

\section{Concluding remarks}
\label{sec:concluding}

In this paper, we have studied the one-dimensional supersymmetric $t-J$
model with the most generic integrable  boundary condition, which is described by the
Hamiltonian (\ref{Hamilton}) and the corresponding integrable boundary terms are
associated with  the most generic non-diagonal $K$-matrices given by
(\ref{K-matrix-1-RA})-(\ref{K+-1}). By combining the  algebraic Bethe ansatz and the
off-diagonal Bethe ansatz, we construct the eigenstates of the transfer matrix in terms of
the nested Bethe states given by (\ref{psi-RA}) and (\ref{state F}), which have well-defined
homogeneous limit. The corresponding eigenvalues are given in terms of the inhomogeneous $T-Q$
relation (\ref{3t-101}) and the associated BAEs (\ref{BAE-Open-3-1})-(\ref{BAE-Open-3-2}).
The exact solution of this paper provides basis for further analyzing the thermodynamic
properties and correlation functions of the model. These are under investigation
and results will be reported elsewhere.

\acknowledgments
We would like to thank Prof. Y. Wang for his valuable discussions and continuous encouragements.
The financial supports from the National Program
for Basic Research of MOST (Grant No. 2016YFA0300600 and
2016YFA0302104), the National Natural Science Foundation of China
(Grant Nos.  11434013, 11425522 and 11547045), the Major Basic Research Program of Natural Science of Shaanxi Province
(Grant No. 2017ZDJC-32), BCMIIS and the Strategic Priority Research Program of the Chinese
Academy of Sciences are gratefully acknowledged.

\bibliographystyle{JHEP}
\bibliography{references}

\end{document}